\documentclass[aps,pre,twocolumn,groupedaddress]{revtex4-1}
\usepackage{graphicx}
\usepackage{dcolumn}
\usepackage{bm}
\usepackage{amsmath}
\usepackage{amssymb}
\usepackage{color}

\begin{document}

%\preprint{}

%Title of paper
\title{Virtual bending method to calculate bending rigidity, saddle-splay modulus, and spontaneous curvature of thin fluid membrane}

\author{Hiroshi Noguchi}
\email[]{noguchi@issp.u-tokyo.ac.jp}
\affiliation{Institute for Solid State Physics, University of Tokyo, Kashiwa, Chiba 277-8581, Japan}

%\date{\today}

\begin{abstract}
A method to calculate the bending rigidity $\kappa$, saddle-splay modulus $\bar{\kappa}$, and spontaneous curvature $C_0$ of a fluid membrane is proposed.
Virtual work for the bending deformations into cylindrical and spherical shapes is calculated for a flat membrane.
This method does not require a force decomposition, unlike the existing stress-profile method.
The first derivative of the deformation 
 gives $\kappa C_0$ and is a discrete form of the first moment of the stress profile.
The second derivatives give $\kappa$ and $\bar{\kappa}$, and include the variance terms of the first derivatives, which are not accounted for in the stress-profile method.
This method is examined for a solvent-free meshless membrane model and a dissipative-particle-dynamics two-bead amphiphilic molecular model.
It is concluded that $\kappa$ and $\bar{\kappa}$ of a thin membrane can be accurately calculated,
whereas for a thick membrane or one with an explicit solvent, a further extension to include the volume-fluctuation effects is required for an accurate estimation.
The amplitude of the volume-fluctuation effects can be evaluated using the parameter dependence in the present method.
\end{abstract}

\maketitle

\section{Introduction}

Amphiphilic molecules self-assemble into various structures including spherical and worm-like micelles, vesicles, and bicontinuous phases~\cite{isra11,safr94}.
Among them, a bilayer membranes have been intensively studied because they are a basic structure of biomembranes.
Biomembranes are typically in a fluid phase,
and their shapes are regulated by many types of proteins via  bending deformation~\cite{mcma05,shib09,baum11,suet14,joha15}.
Because lipid membranes maintain an almost constant area,
the bending deformation is the most important factor in understanding the biomembrane morphology.

The free energy of a fluid membrane is given by
\begin{equation} \label{eq:helcv}
F = \int \left[\frac{\kappa}{2}\left(C_1+C_2 -C_0  \right)^2 + \bar{\kappa} C_1C_2 + \gamma_{\rm it} \right] dA,
\end{equation} 
where $C_1$ and $C_2$ are the principal curvatures, and it is integrated over a membrane surface with an area of $A$~\cite{safr94,canh70,helf73}.
This denotes the energy expansion of the curvatures to the second order, with coefficients of
the bending rigidity $\kappa$, saddle-splay modulus or Gaussian curvature modulus $\bar{\kappa}$, and
spontaneous curvature $C_0$. 
$\gamma_{\rm it}$ is the internal (bare) surface tension conjugated to the real membrane area $A$~\cite{davi91,fara03a,shib16,gueg17}.
The saddle-splay modulus $\bar{\kappa}$ does not contribute to the shape transformation of a vesicle
with a fixed topology,
because of the Gauss--Bonnet theorem, $\oint C_1C_2 dA= 4\pi (1-g)$, where $g$ is the genus of the vesicle.

The bending rigidity $\kappa$ can be estimated by several methods based on the fluctuation analysis or mechanical response.
The fluctuation spectrum of a flat membrane, $\langle |h(q)|^2\rangle=k_{\rm B}T/(\gamma q^2 + \kappa q^4)$,
is the most widely used to calculate $\kappa$ in simulations~\cite{safr94,goet99,lind00,shib11}.
$\gamma$ is the mechanical (frame) surface tension conjugated to the projected membrane area $A_{xy}=L_xL_y$, 
where $L_x$ and $L_y$ are the side lengths of the simulation box and the membrane is normal to the $z$ direction~\cite{shib16}.
Moreover, $\kappa$ can be calculated from the fluctuation spectra of spherical~\cite{schn84,helf86,miln87,gomp04c,dimo14} 
and cylindrical~\cite{zhon89,four07,barb09,shib11} membranes.
For a lipid bilayer,  $\kappa$ can also be calculated from fluctuations in lipid orientation~\cite{wats12}.
Experimentally,  $\kappa$ is the most commonly measured by a tubular (tether) membrane extension from a vesicle
using the force strength and surface tension~\cite{dimo14,bo89,evan96,cuve05}. 
In simulations,  $\kappa$ can be calculated from the force strength and radius of a cylindrical membrane~\cite{harm06,shib11}.
Moreover, $\kappa$ can also be calculated from the anisotropic lateral stress of a buckled membrane in simulations~\cite{nogu11a,hu13a}.

The spontaneous curvature $C_0$ can be calculated from the force dependence on the radius of a cylindrical membrane~\cite{shib11}
and the curvature of a membrane strip~\cite{shib11,mahm19}.
The saddle-splay modulus $\bar{\kappa}$ can be calculated from the shape transition between a flat membrane patch and vesicle~\cite{hu12}.
For a membrane with a constant spontaneous curvature, $\bar{\kappa}$ can also be calculated from the curvature of a membrane patch~\cite{nogu19}.

The aforementioned methods typically require a large membrane size.
Helfrich~\cite{helf81} and Safran and coworkers~\cite{safr94,szle90} developed 
calculation formulas for $\kappa C_0$ and $\bar{\kappa}$ from the first and second moments of the stress profile of a flat membrane, respectively.
These formulas require a relatively small membrane size.
However, the force decomposition of multibody potentials is required to calculate the stress profile.
Three- or four-body potentials can be uniquely decomposed into pairwise central forces (central force decomposition (CFD)) 
in the three-dimensional (3D) space~\cite{adma10}, 
although other decompositions are also available for three-body potentials (force-center decomposition (FCD) and their hybrids)~\cite{naka16}.
Recently, a covariant CFD was proposed for the unique decomposition of $n$-body forces at $n>4$~\cite{torr15,torr16}.
However, it cannot properly decompose the forces caused by several pairwise potentials, as described in Appendix~\ref{sec:ccfd}.
The obtained value of $\bar{\kappa}$ is largely dependent on the force decomposition~\cite{naka16},
and deviates from the expected value even in a solvent-free molecular model consisting only of pairwise potentials~\cite{hu12}.
In principle, macroscopic quantities such as  $\bar{\kappa}$ should not depend on an arbitrary choice of local quantities.
The first moment of the stress profile has been investigated  significantly less because $C_0=0$ for a symmetric membrane.
To our knowledge, $\kappa C_0$ was calculated using the stress-profile method for the monolayer consisting of a bilayer of an solvent-free molecular model~\cite{hu12}
and for monolayers in the solvent interface and bilayer using a self-consistent field theory~\cite{ting17}.

In this study, we propose a calculation method for $\kappa$, $\bar{\kappa}$, and $\kappa C_0$ from a flat membrane with a relatively small computational cost.
Virtual bending deformations are considered, and the free-energy change is  directly calculated from the forces and second derivatives of potentials
without using force decompositions.
Although a similar idea was previously considered by Farago and Pincus~\cite{fara04},
they used incorrect deformations; a shear deformation (tilt for lipids) is involved in area expansion, and the lateral deformations are missing in bending deformations.
In Sec.~\ref{sec:theory}, we first outline the previous method based on the stress profile
and then propose the virtual bending method.
We show that the first derivative of the virtual bending method is the discrete form of  the first moment of the stress profile 
in Sec.~\ref{sec:first}.
We apply this method to a solvent-free meshless membrane and an explicit-solvent bilayer membrane in Sec.~\ref{sec:meshless} and \ref{sec:DPD}, respectively.
It works excellently for the former, but there is scope for further improvement in the volume-fluctuation effects in the latter case.
A summary and discussion are given in Sec.~\ref{sec:sum}.

\section{Theory}\label{sec:theory}

\subsection{Previous Method Based on the Stress Profile}\label{sec:stpro}

First, we present the formulas to calculate $\kappa C_0$ and $\bar{\kappa}$ from the stress profile~\cite{helf81,szle90}.
The bending deformation is divided into lateral and vertical deformations using a pyramid approximation,
in which the normal direction is fixed~\cite{szle90}.
In the lateral deformation, the membrane thickness is fixed, and
the local volume change of the membrane is given by
\begin{equation} \label{eq:lv}
\delta V = A_0(2z dz \delta H +z^2 dz \delta K),
\end{equation}
where $\delta H$ and $\delta K$ are the local variations of the mean curvature $H=(C_1+C_2)/2$ 
and Gaussian curvature $K=C_1C_2$, respectively.
The origin of the vertical ($z$) coordinate is set to maintain
a constant lateral area $A_0$ at $z=0$ under the bending deformation.
In the vertical deformation,
the membrane thickness changes to maintain its volume.
The work done per unit area to make these deformations is
separately calculated for lateral and vertical deformations.
Using the expansion of the stresses to the first order of $H$ and comparing with Eq.~(\ref{eq:helcv}),
the first and second moments of the stress profile give
$\kappa C_0$ and $\bar{\kappa}$, respectively:
\begin{eqnarray} \label{eq:stc0}
-\kappa C_0 &=& \int (\sigma_{\parallel} (z) - \sigma_{zz}(z)) z\ dz, \\ \label{eq:stkb}
\bar{\kappa} &=&  \int (\sigma_{\parallel} (z) - \sigma_{zz}(z)) z^2\ dz,
\end{eqnarray} 
where $\sigma_{\parallel} (z)$ and $\sigma_{zz}(z)$ are the lateral and vertical stresses, respectively.

\subsection{Virtual Work for Deformation}\label{sec:vw}

For a perturbation variable $\lambda$ under constant volume and temperature,
the first and second derivatives of the free energy $F$ are derived as
\begin{eqnarray} \label{eq:free}
F &=& - k_{\rm B}T \ln \left(\int e^{-E/k_{\rm B}T}\ dE\right),\\ \label{eq:fl1}
  \left.\frac{\partial F}{\partial \lambda}\right|_{V,T} &=& \frac{\int \frac{\partial E}{\partial \lambda} e^{-E/k_{\rm B}T} dE}{\int e^{-E/k_{\rm B}T} dE} 
= \Big\langle   \frac{\partial E}{\partial \lambda} \Big\rangle_{V,T},\\ \label{eq:fl2}
 \left.\frac{\partial^2 F}{\partial \lambda^2}\right|_{V,T} &=&  \Big\langle   \frac{\partial^2 E}{\partial \lambda^2} \Big\rangle_{V,T} \\ \nonumber
&& - \frac{1}{k_{\rm B}T}\Big(  \Big\langle \Big(\frac{\partial E}{\partial \lambda}\Big)^2 \Big\rangle_{V,T} 
- \Big\langle \frac{\partial E}{\partial \lambda} \Big\rangle_{V,T}^2  \Big),
\end{eqnarray} 
where  $E$ and  $k_{\rm B}T$ are the internal and thermal energies, respectively, and
$\langle ... \rangle$  represents the ensemble average.
The first derivative of the free energy is the ensemble average of the derivative of the internal energy.
In contrast, the second derivative also has a second (variance) term, which indicates the thermal-fluctuation effects.
For bending deformations,
a curvature $C$ is taken as a perturbation variable with constant membrane area: $\lambda=C$.

In the derivation of the previous method using the stress profile [Eq.~(\ref{eq:stkb})], this second term is not considered.
In typical molecular simulations, the thermal fluctuations are significant and thus,
the second term is non-negligible.
Therefore, the stress-profile method is not applicable to molecular simulations.
In particular, for  lipid membranes, molecular fluctuations and entropy are key factors in distinguishing
a fluid phase and gel or crystal phases.

\subsection{Surface Tension}\label{sec:ten}

Let us consider that a flat fluid membrane consists of $N$ particles
(atoms or coarse-grained particles representing multiple atoms), which
 interact with each other via a potential $U({\bf r}_1,...,{\bf r}_N)$.
The membrane is connected by periodic boundary conditions in the $xy$ plane.

Before discussing the bending deformation,
we consider the affine deformation to calculate the mechanical surface tension $\gamma$ as a simpler system.
The affine deformation for an isotropic surface expansion in the $xy$ plane is expressed as
\begin{equation} \label{eq:affine}
{\bf r}'_i= 
\left(
\begin{array}{ccc}
1+\varepsilon/2 & 0 & 0 \\ 
0 & 1+\varepsilon/2 & 0 \\
0 & 0 & 1-\varepsilon
\end{array}
 \right) {\bf r}_i,
\end{equation}
where ${\bf r}_i$ and ${\bf r}'_i$ are 
the positions of the $i$-th particle before and after the deformation, respectively.
The volume is fixed in $O(\varepsilon)$ for a small deformation of $\varepsilon\ll 1$.

The free-energy change due to this deformation is caused by the surface energy
$\Delta F = \gamma \Delta A_{xy}$.
Therefore, the surface tension is given by
\begin{equation} \label{eq:dfaf}
\gamma = \frac{1}{A_{xy}} \sum_i \Big\langle \frac{1}{2}\Big(x_i\frac{\partial U}{\partial x_i} + y_i\frac{\partial U}{\partial y_i} \Big) - z_i\frac{\partial U}{\partial z_i}\Big\rangle ,
\end{equation}
using $\Delta A_{xy}= \varepsilon A_{xy}$ with Eqs.~(\ref{eq:fl1}) and (\ref{eq:affine}).
This expression is the same as derived from the virial expression of the stress:
$\gamma = [(\sigma_{xx}+\sigma_{yy})/2 - \sigma_{zz}]/A_{xy}$.
Because the fluid membrane is laterally isotropic,  $\sigma_{xx}=\sigma_{yy}=\sigma_{\parallel}$.
To obtain a better accuracy, the lateral stress is usually calculated as the average of two stresses, 
$\sigma_{\parallel}=(\sigma_{xx}+\sigma_{yy})/2$ in molecular simulations.
Moreover, in the numerical calculation of Eq.~(\ref{eq:dfaf}), 
 the different origin of the coordinate should be employed
for each potential to reduce numerical errors as described in Ref.~\cite{naka16}.
When the potential interaction crosses the periodic boundary,
the periodic image ($x_i \pm x_xL_x$, $y_i \pm n_yL_y$) is employed, as in the force calculation.

Thus, the expression for the surface tension can be derived from the affine deformation.
Note that the local stress decomposition is not required for this calculation,
so it is free from the non-uniqueness of the local stress field.

\begin{figure}
\includegraphics{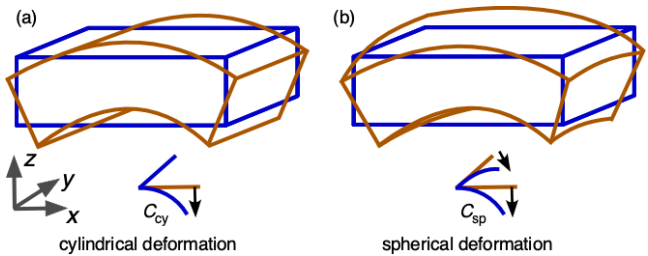}
\caption{
Schematic drawing of the bending deformations of a fluid membrane.
(a) Cylindrical deformation. The membrane bends along the $x$ direction with a curvature of $C_{\rm cy}$.
(b) Spherical deformation.  The membrane bends isotropically in the $xy$ plane with a curvature of $C_{\rm sp}$.
}
\label{fig:sch}
\end{figure}

\subsection{Virtual Bending Method}\label{sec:vb}

We consider two types of bending deformation of a flat fluid membrane: 
deformation into cylindrical and spherical shapes, as depicted in Fig.~\ref{fig:sch}.
One cannot globally bend the membrane with a constant curvature under periodic boundary conditions.
However, the bending deformation of a local region is possible.
We consider the local deformations as described below in this subsection.
Here, only local potential interactions are considered; that is,
the maximum interaction distances of the potentials are shorter than 
half the minimum side length of the simulation box.

As in the affine deformation described in Sec.~\ref{sec:ten},
the local volume is fixed during the deformation.
The origin of the $z$ coordinate is set to maintain a constant
 lateral area at $z=0$ under the cylindrical deformation, as considered in Sec.~\ref{sec:stpro}.
This plane of $z=0$ deforms into a cylindrical arc or spherical cap with a curvature radius of  $1/C_{\rm cy}$ or
 $1/C_{\rm sp}$, respectively.

For the  cylindrical and spherical deformations,
the free energy is expressed as $F=(\kappa/2)(C_{\rm cy} - C_0)^2A_{xy}$ and 
$F=[(\kappa/2)(2C_{\rm sp} - C_0)^2 + \bar{\kappa}C_{\rm sp}^2]A_{xy}$, respectively.
Therefore, the first and second derivatives are given by
\begin{eqnarray} \label{eq:cylf1}
 \left.\frac{\partial F}{\partial C_{\rm cy}}\right|_{V,T,C_{\rm cy}=0} &=& -\kappa C_0A_{xy}
= \Big\langle   \frac{\partial U}{\partial C_{\rm cy}} \Big\rangle,\\ \label{eq:cylf2}
 \left.\frac{\partial^2 F}{\partial C_{\rm cy}^2}\right|_{V,T,C_{\rm cy}=0} 
 &=& \kappa A_{xy}
=  \Big\langle   \frac{\partial^2 U}{\partial C_{\rm cy}^2} \Big\rangle \\ \nonumber
&& - \frac{1}{k_{\rm B}T}\Big(  \Big\langle \Big(\frac{\partial U}{\partial C_{\rm cy}}\Big)^2 \Big\rangle - \Big\langle \frac{\partial U}{\partial C_{\rm cy}} \Big\rangle^2  \Big), \\ \label{eq:sphf2}
\left.\frac{\partial^2 F}{\partial C_{\rm sp}^2}\right|_{V,T,C_{\rm sp}=0} 
 &=& (4\kappa+2\bar{\kappa}) A_{xy} 
=  \Big\langle   \frac{\partial^2 U}{\partial C_{\rm sp}^2} \Big\rangle \\ \nonumber
&& - \frac{1}{k_{\rm B}T}\Big(  \Big\langle \Big(\frac{\partial U}{\partial C_{\rm sp}}\Big)^2 \Big\rangle - \Big\langle \frac{\partial U}{\partial C_{\rm sp}} \Big\rangle^2  \Big),
\end{eqnarray}
where the averages are taken at a constant volume and temperature of the flat membrane.
The bending rigidity $\kappa$ is calculated using Eq.~(\ref{eq:cylf2}),
and the saddle-splay modulus  $\bar{\kappa}$ is calculated from  Eqs.~(\ref{eq:cylf2}) and (\ref{eq:sphf2})
as $\bar{\kappa}=(\partial^2 F/\partial C_{\rm sp}^2|_{V,T,C_{\rm sp}=0}/2 - 2\partial^2 F/\partial C_{\rm cy}^2|_{V,T,C_{\rm cy}=0})/A_{xy}$.
The last terms of Eqs.~(\ref{eq:cylf2}) and (\ref{eq:sphf2}) on the fluctuations are missing in the previous method described in Sec.~\ref{sec:stpro}.
Note that $\partial F/\partial C_{\rm sp}|_{V,T,C_{\rm sp}=0}$ does not give additional information, since
$\partial F/\partial C_{\rm sp}|_{V,T,C_{\rm sp}=0} = - 2\kappa C_0A_{xy} = 2 \partial F/\partial C_{\rm cy}|_{V,T,C_{\rm sp}=0}$.

To calculate $\langle\partial U/\partial C_{\rm cy}\rangle$, $\langle\partial^2 U/\partial C_{\rm cy}^2\rangle$, and $\langle\partial^2 U/\partial C_{\rm sp}^2\rangle$,
 local deformations of the membrane are considered.
In molecular simulations, the potential $U$ typically consists of multiple types of local potentials as
\begin{equation}
U({\bf r}_1,...,{\bf r}_N)= \sum_{n=2}^{N} \sum_{k_n} U_{k_n}({\bf r}_{k_n,1},...,{\bf r}_{k_n,n}),
\end{equation}
where each $U_{k_n}$ is an $n$-body potential that is invariant under translation and rotation.
The energy change of each potential $U_{k_n}$ is determined by
the relative positional changes of the interacting particles ${\bf r}_{k_n,1},...,{\bf r}_{k_n,n}$.
The magnitudes of these positional changes in the second-order approximation 
depend on the origin ${\bf r}_{\rm cc}$ of the deformation
in the $xy$ plane, unlike in the first-order level.
For pairwise potentials, the middle position of two interacting particles ($i$ and $j$) is
taken from the geometrical symmetry as ${\bf r}_{\rm cc}= ({\bf r}_i+{\bf r}_j)/2$.
For multibody potentials,
the geometrical center,  ${\bf r}_{\rm cc}= \sum_j^n {\bf r}_j/n$, can be the  center position;
however, other choices are also available. 
We have examined different choices for the meshless membrane
and observe no significant effects on the estimated values 
when the center is chosen in a reasonable manner,
as described in the next section.

After the cylindrical deformation, 
the position of the $i$-th particle is given by
\begin{eqnarray} \label{eq:cyl0}
x'_i &=& r'_i \sin \theta_{{\rm c},i} , \\
y'_i &=& y_i ,  \\
z'_i &=& r'_i \cos \theta_{{\rm c},i} - \frac{1}{C_{\rm cy}},
\end{eqnarray}
where  
\begin{eqnarray} \label{eq:cylth}
\theta_{{\rm c},i} &=& C_{\rm cy} x_i, \\  \label{eq:cylr}
r'_{{\rm c},i} &=& \frac{1}{C_{\rm cy}} + z_{i} - \frac{1}{2}z_i^2 C_{\rm cy} + \frac{\alpha_1}{2}z_i^3 C_{\rm cy}^2.
\end{eqnarray}
The local volume is kept constant at $\alpha_1=1$ in the order of $C_{\rm cy}^2$.
For the analogy of the second term in the free-energy expansion (Eq.~(\ref{eq:fl2})),
the volume-fluctuation term 
might be required for the second order of $C_{\rm cy}$ 
as a function of the variance of the local volume.
Although the total volume is fixed, the local volume fluctuates. 
To examine its effects, the factor $\alpha_1$ is considered here.
Similarly, the cylindrical deformation in the $y$ direction is obtained.

Using the expansion of $\cos \theta$ and $\sin \theta$ to the second order,
the positional change is rigorously expressed in the second order of $C_{\rm cy}$ as
\begin{eqnarray} \label{eq:cylx}
x'_i &=&  x_i \left[ 1 + z_i C_{\rm cy} - \left(\frac{z_i^2}{2} + \frac{x_i^2}{6} \right) C_{\rm cy}^2\right] + O(C_{\rm cy}^3), \\ 
z'_i &=& z_i - \frac{1}{2}(z_i^2 + x_i^2) C_{\rm cy} +  \frac{z_i}{2}(\alpha_1 z_i^2 - x_i^2) C_{\rm cy}^2 + O(C_{\rm cy}^3). \nonumber \\ \label{eq:cylz}
\end{eqnarray}

Similarly, for the spherical deformation, 
the positional change is expressed in the second order of $C_{\rm sp}$ as
\begin{eqnarray} \label{eq:sphx}
{\rho}'_i &=&  r'_{{\rm s},i} \sin \theta_{{\rm s},i}  \\ \nonumber 
&=&  \rho_i \left[ 1 + z_i C_{\rm sp} - \left(z_i^2 + \frac{\rho_i^2}{6} \right) C_{\rm sp}^2\right] + O(C_{\rm sp}^3), \\ \label{eq:sphz}
z'_i &=& r'_{{\rm s},i} \cos \theta_{{\rm s},i} - \frac{1}{C_{\rm sp}} \\ \nonumber 
&=& z_i - \left(z_i^2 + \frac{\rho_i^2}{2} \right) C_{\rm sp} +  z_i\left( \frac{5\alpha_2}{3}z_i^2 -  \frac{\rho_i^2}{2}\right) C_{\rm sp}^2 + O(C_{\rm sp}^3).
\end{eqnarray}
where ${\rho}_i= \sqrt{x_i^2+y_i^2}$, ${\rho}'_i= \sqrt{{x'}_i^2+{y'}_i^2}$, and  $\theta_{{\rm s},i} = C_{\rm sp} \rho_i$.
The position is radially varied as
\begin{equation} \label{eq:sphr}
r'_{{\rm s},i} = \frac{1}{C_{\rm sp}} + z_{i} - z_i^2 C_{\rm sp} + \frac{5\alpha_2}{3}z_i^3 C_{\rm sp}^2,
\end{equation}
where the local volume is kept constant at $\alpha_2=1$.

The energy change of the potential $U_{k_n}$ for a small deformation is given by
\begin{eqnarray} \label{eq:du0}
\delta U_{k_n} &\simeq&  \sum_{j}^{n} \left(\delta x_j \frac{\partial U_{k_n}}{\partial x_j} + \delta y_j \frac{\partial U_{k_n}}{\partial y_j} + \delta z_j \frac{\partial U_{k_n}}{\partial z_j}\right) \\ \nonumber
&& + \frac{1}{2}\left(\sum_{j}^{n} \delta x_j \frac{\partial }{\partial x_j} + \delta y_j \frac{\partial }{\partial y_j} + \delta z_j \frac{\partial }{\partial z_j}\right)^2 U_{k_n}.
\end{eqnarray}
For the cylindrical deformation,
 $\delta U_{k_n}$ is expressed as a function of $C_{\rm cy}$  by substituting $\delta x_j=x'_j-x_j$, $\delta y_j=0$, 
and $\delta z_j=z'_j-z_j$ of Eqs.~(\ref{eq:cylx}) and (\ref{eq:cylz}) into Eq.~(\ref{eq:du0}).
Subsequently, after averaging $\delta U_{k_n}$ for the cylindrical deformations along the $x$ and $y$ directions,
 the first and second derivatives are obtained as
\begin{eqnarray} \label{eq:cyluc1}
  \frac{\partial U_{k_n}}{\partial C_{\rm cy}}  &=& 
\sum_{j}^{n} \frac{z_j}{2}\left(x_j \frac{\partial U_{k_n}}{\partial x_j} + y_j \frac{\partial U_{k_n}}{\partial y_j}\right) \\ \nonumber
&& - \left( \frac{z_j^2}{2} +  \frac{x_j^2+y_j^2}{4}  \right)\frac{\partial U_{k_n}}{\partial z_j}, \\ \label{eq:cyluc2}
  \frac{\partial^2 U_{k_n}}{\partial C_{\rm cy}^2} &=&  
\sum_{j}^{n}  \Bigg\{ - x_j\left(\frac{z_j^2}{2} + \frac{x_j^2}{6} \right)\frac{\partial U_{k_n}}{\partial x_j}  \\ \nonumber
&&                       - y_j\left(\frac{z_j^2}{2} + \frac{y_j^2}{6} \right)\frac{\partial U_{k_n}}{\partial y_j} \\ \nonumber
&&                       + z_j\left(\alpha_1 z_j^2 - \frac{x_j^2+y_j^2}{2} \right)\frac{\partial U_{k_n}}{\partial z_j}  \Bigg\} \\ \nonumber
&& + \frac{1}{2} \Bigg\{ \left(\sum_{j}^{n} x_jz_j\frac{\partial }{\partial x_j} - \frac{z_j^2+x_j^2}{2}\frac{\partial }{\partial z_j} \right)^2 U_{k_n} \\ \nonumber
&& + \left(\sum_{j}^{n} y_jz_j\frac{\partial }{\partial y_j} - \frac{z_j^2+y_j^2}{2}\frac{\partial }{\partial z_j} \right)^2 U_{k_n} \Bigg\}.
\end{eqnarray}
Similarly, for the spherical deformation, substituting $\delta x_j/x_j=\delta y_j/y_j= (\rho'_j-\rho_j)/\rho_j$
and $\delta z_j=z'_j-z_j$ of Eqs.~(\ref{eq:sphx}) and (\ref{eq:sphz}) into Eq.~(\ref{eq:du0}) yields
\begin{eqnarray}
 \frac{\partial^2 U_{k_n}}{\partial C_{\rm sp}^2} &=& 
\sum_{j}^{n}  \Bigg\{- \Big(2z_j^2 + \frac{x_j^2+y_j^2}{3} \Big)
\Big(x_j\frac{\partial U_{k_n}}{\partial x_j} + y_j\frac{\partial U_{k_n}}{\partial y_j}\Big)  \nonumber \\  \nonumber 
&&                       + z_j\left[\frac{10}{3} \alpha_2 z_j^2 - (x_j^2+y_j^2) \right]\frac{\partial U_{k_n}}{\partial z_j}  \Bigg\}   \\ \nonumber
&& +  \Bigg\{ \bigg[\sum_{j}^{n} x_jz_j\frac{\partial }{\partial x_j} + y_jz_j\frac{\partial }{\partial y_j} \\
&& - \Big(z_j^2+\frac{x_j^2+y_j^2}{2}\Big)\frac{\partial }{\partial z_j} \bigg]^2 U_{k_n} \Bigg\}. \label{eq:sphc2}
\end{eqnarray}
As expected, $\partial U_{k_n}/\partial C_{\rm sp} = 2\partial U_{k_n}/\partial C_{\rm cy}$ is obtained. 

For a pairwise potential $U_{\rm pair}(r_{ij})$ between the $i$-th and $j$-th particles,
\begin{eqnarray} \label{eq:pair1}
 \frac{\partial U_{\rm pair}}{\partial C_{\rm cy}} &=& 
 \Big(\frac{\rho_{ij}^2}{2} - z_{ij}^2\Big) \frac{z_{\rm G}}{r_{ij}}\frac{\partial U_{\rm pair}}{\partial r_{ij}}, \\ \label{eq:pair2}
 \frac{\partial^2 U_{\rm pair}}{\partial C_{\rm cy}^2} &=&  
 \bigg[ (3\alpha_1+1)z_{ij}^2z_{\rm G}^2 + \frac{\alpha_1}{4}z_{ij}^4  \\ \nonumber
&& - \frac{x_{ij}^4+y_{ij}^4}{24} - \frac{\rho_{ij}^2z_{ij}^2}{4}\bigg]\frac{1}{r_{ij}}\frac{\partial U_{\rm pair}}{\partial r_{ij}} \\ \nonumber
&& + \frac{z_{\rm G}^2}{2}\big[(x_{ij}^2-z_{ij}^2)^2+(y_{ij}^2-z_{ij}^2)^2\big] \\ \nonumber
&& \Big(\frac{1}{r_{ij}^2}\frac{\partial^2 U_{\rm pair}}{\partial r_{ij}^2}
-\frac{1}{r_{ij}^3}\frac{\partial U_{\rm pair}}{\partial r_{ij}}\Big), \\ \label{eq:pair3}
 \frac{\partial^2 U_{\rm pair}}{\partial C_{\rm sp}^2} &=&  
 \bigg[ (10\alpha_2+4)z_{ij}^2z_{\rm G}^2  -\rho_{ij}^2z_{\rm G}^2  \\ \nonumber
&& - \frac{3\rho_{ij}^2z_{ij}^2}{4} - \frac{\rho_{ij}^4}{12} + \frac{5\alpha_2 z_{ij}^4}{6} \bigg] \frac{1}{r_{ij}}\frac{\partial U_{\rm pair}}{\partial r_{ij}} \\ \nonumber
&& + \big( \rho_{ij}^4 + 4z_{ij}^4 - 4\rho_{ij}^2z_{ij}^2\big)z_{\rm G}^2 \\ \nonumber
&& \Big(\frac{1}{r_{ij}^2}\frac{\partial^2 U_{\rm pair}}{\partial r_{ij}^2}-\frac{1}{r_{ij}^3}\frac{\partial U_{\rm pair}}{\partial r_{ij}}\Big), 
\end{eqnarray}
where $r_{ij}=|{\bf r}_{ij}|$, ${\bf r}_{ij}={\bf r}_{i}-{\bf r}_{j}$, ${\bf r}_{\rm G}\equiv (x_{\rm G},y_{\rm G},z_{\rm G}) = ({\bf r}_i+{\bf r}_j)/2$
 and $\rho_{ij}^2=x_{ij}^2+y_{ij}^2$.
Equations~(\ref{eq:pair2}) and (\ref{eq:pair3}) are derived using the origin ${\bf r}_{\rm cc}= {\bf r}_{\rm G}$,
whereas this is not necessary for the derivation of Eq.~(\ref{eq:pair1}).

\subsection{First Derivatives and First Moments of Stress Profiles}\label{sec:first}

The first derivative of the cylindrical deformation [Eqs.~(\ref{eq:cylf1}) and (\ref{eq:cyluc1})]
gives  $-\kappa C_0$  by dividing by $A_{xy}$.
This coincides with the formula for $\kappa C_0$ using the first moment of the stress-profile method [Eq.~(\ref{eq:stc0})]
and is independent of the choice of the origin ${\bf r}_{\rm cc}$ of the deformation in the $xy$ plane.
For pairwise potentials, this relation  is straightforwardly determined from  Eq.~(\ref{eq:pair1}).
For multibody potentials, it is derived as follows.
An $n$-body potential $U_{k_n}$ can be expressed as a function of the distances of $n(n-1)/2$ particle pairs 
as $U_{k_n}=\breve{U}_{k_n}({\bf R})$, where ${\bf R}$ is an ($n(n-1)/2$)-dimensional vector of the particle-pair distances~\cite{adma10,adma11}.
For $n>3$ and $n>4$, $\breve{U}_{k_n}({\bf R})$ is not uniquely determined in the 2D and 3D spaces, respectively.
The force decomposition into central forces (CFD) is expressed as 
$\partial U_{k_n}/\partial {\bf r}_i= -\sum_j f_{ij}\hat{\bf r}_{ij}$,
where $f_{ij}= -\partial \breve{U}_{k_n}/\partial r_{ij}$ and  $\hat{\bf r}_{ij}= {\bf r}_{ij}/r_{ij}$.
Hence, the first derivative $\langle \partial U_{k_n}/\partial C_{\rm cy}\rangle$ 
is given by the sum of the stress contributions of the central force $f_{ij}$, 
as $\partial U_{\rm pair}/\partial r_{ij}$ is replaced by $-f_{ij}$ in Eq.~(\ref{eq:pair1}).
This relation is valid for any choice of $\breve{U}_{k_n}({\bf R})$.
Therefore, the present method derives the same formula for $\kappa C_0$ as the stress-profile method for 
any potential of $U_{k_n}$.
Conversely, the expression using Eq.~(\ref{eq:cyluc1}) is interpreted as a discrete form of the first moment of the stress profile.
This means that the first moment of the stress profile is not modified by
the choice of CFD.
The FCD and hybrid decomposition for three-body potentials~\cite{naka16} also give the correct value of the first moment.
By substituting Eqs.~(10) and (13) in Ref.~\onlinecite{naka16} to the first moment of FCD, that of CFD is obtained.
By contrast, the Goetz--Lipowsky force decomposition~\cite{goet99}, ${\bf f}_{ij}=({\bf f}_{i}-{\bf f}_{j})/n$, which does not conserve angular momentum,
gives an incorrect value.
Thus, the angular-momentum conservation is likely to be a necessary condition for force decomposition to give the first moment correctly.

In contrast to the first derivative,
the second derivatives depend on the choice of ${\bf r}_{\rm cc}$.
When the forces are decomposed, the second derivatives and second moment of the stress profile are dependent on  $\breve{U}_{k_n}({\bf R})$.

\section{meshless membrane}\label{sec:meshless}
\subsection{Model}

\begin{figure}
\includegraphics{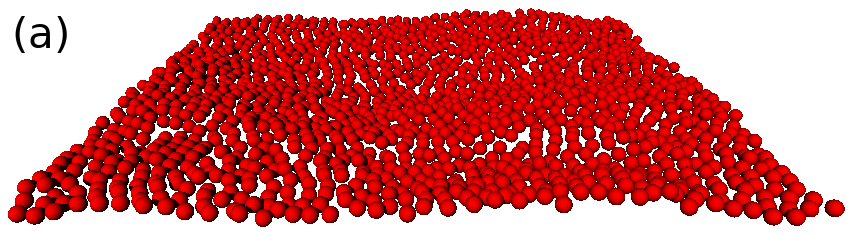}
\includegraphics{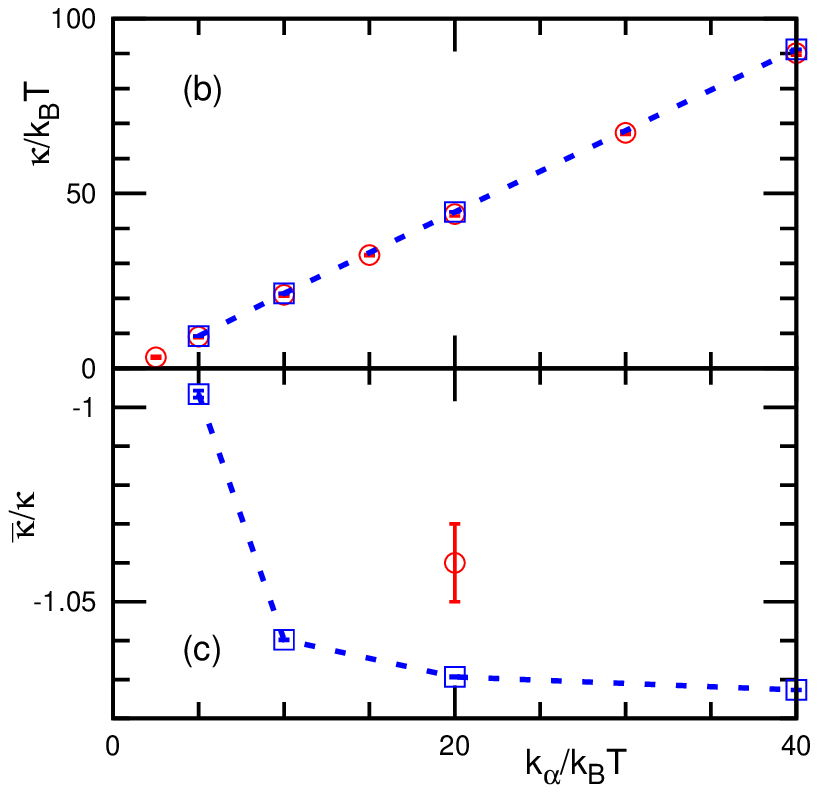}
\caption{
Calculation of the bending rigidity $\kappa$ and saddle-splay modulus $\bar{\kappa}$
of the meshless membrane model.
(a) Snapshot of the membrane at $k_{\alpha}/k_{\rm B}T=10$ and $N=1600$.
(b, c) Dependence of (b) $\kappa$ and (c) $\bar{\kappa}$ on  $k_{\alpha}$.
The squares and dashed lines represent the data calculated from the present method with $\alpha_1=\alpha_2=1$ for $N=100$.
The circles represent the data calculated from (b) the membrane-fluctuation spectrum at $N=1600$ and  
(c) the membrane closing probability described in Appendix~\ref{sec:close}.
}
\label{fig:mls1}
\end{figure}

To examine the virtual bending method,
we first apply it to a solvent-free meshless membrane model.
It was proposed by us in Ref.~\onlinecite{nogu06} and applied to various problems, including self-assembly dynamics, membrane rupture~\cite{nogu06a}, 
membrane buckling~\cite{nogu11a}, and the interaction of the binding sites between membranes~\cite{nogu13}.
The membrane is represented by a self-assembled one-layer sheet of $N$ particles,
as shown in Fig.~\ref{fig:mls1}(a).
The details of this meshless membrane model are described in Ref.~\onlinecite{nogu06};
we briefly explain it here.

The particles interact with each other via the potential
$U= \varepsilon( \sum_{i<j} U_{{\rm rep},ij} +  \sum_{i} U_{{\rm att},i}) + k_{\alpha}\sum_{i} U_{{\rm \alpha},i}$,
which consists of a soft-core excluded-volume potential $U_{{\rm rep},ij}$ with a 
diameter $\sigma$, an attractive potential $U_{{\rm att},i}$, and a 
curvature potential $U_{{\rm \alpha},i}$.
The excluded-volume potential is given by a pairwise potential,
$U_{{\rm rep},ij}= \exp[-20(r_{ij}/\sigma-1)+0.126]f_{\rm {cut}}(r_{ij}/\sigma)$.
The interaction is smoothly cutoff by a $C^{\infty}$ cutoff function \cite{nogu06}: 
\begin{equation}
f_{\rm {cut}}(s)=
\exp\bigg[A\Big(1+\frac{1}{(|s|/s_{\rm {cut}})^{12} -1}\Big)\bigg]\theta(s_{\rm {cut}}-s),
\end{equation}
where $\theta(s)$ denotes the unit step function.
For $U_{{\rm rep},ij}$, the parameters $A=1$ and $s_{\rm {cut}}=1.2$ are used.

The potential $U_{{\rm att},i}$ is a function of 
 the local density of particles 
$\rho_i= \sum_{j} f_{\rm {cut}}(r_{ij}/\sigma)$,
with the parameters $s_{\rm {half}}=1.8$, $s_{\rm {cut}}=2.1$,
and  $A=\ln(2) [(s_{\rm {cut}}/s_{\rm {half}})^{12}-1]$.
Here, $\rho_{i}$ denotes the number of particles in a 
sphere of radius approximately 
$r_{\rm {att}} = s_{\rm {half}}\sigma$. The potential $U_{{\rm att},i}$ 
is given by
$U_{{\rm att},i} = 0.25\ln[1+\exp\{-4(\rho_i-\rho^*)\}]- C$
where $C= 0.25\ln\{1+\exp(4\rho^*)\}$.
This multibody potential acts as  a pair potential 
$U_{{\rm att},i}\simeq -\rho_i$ with the cutoff at $\rho_i\simeq \rho^*$, 
and it can stabilize
the fluid phase of membranes over a wide range of parameter sets.

The curvature potential is given by 
$U_{{\rm \alpha},i}= \alpha_{\rm {pl}}({\bf r}_{i})$.
The shape parameter aplanarity $\alpha_{\rm {pl}}$ is defined as
\begin{equation}
\alpha_{\rm {pl}}
= \frac{9\lambda_1\lambda_2\lambda_3}{(\lambda_1+\lambda_2+\lambda_3)
    (\lambda_1\lambda_2+\lambda_2\lambda_3+\lambda_3\lambda_1)},
\end{equation}
where $\lambda_1$, $\lambda_2$, and $\lambda_3$ are
the eigenvalues of the weighted gyration tensor,
$a_{\alpha\beta}= \sum_j (\alpha_{j}-\alpha_{\rm G})
(\beta_{j}-\beta_{\rm G})w_{\rm {mls}}(r_{ij})$,
where $\alpha, \beta=x,y,z$ and ${\bf r}_{\rm G}=\sum_j {\bf r}_{j}w_{\rm {mls}}(r_{ij})/\sum_j w_{\rm {mls}}(r_{ij})$.
The aplanarity $\alpha_{\rm {pl}}$ 
represents the degree of deviation from a plane, and it is proportional to $\lambda_1$
 for $\lambda_1 \ll \lambda_2, \lambda_3$.
A Gaussian function with a $C^{\infty}$ cutoff~\cite{nogu06} 
is employed as a weight function: 
\begin{equation}
w_{\rm {mls}}(r_{ij})=
\exp \Big[\frac{(r_{ij}/1.5\sigma)^2}{(r_{ij}/3\sigma)^{12} -1}\Big]\theta(3\sigma-r_{ij}).
\end{equation}

In this study, we use $N=100$ or $1600$, $\varepsilon/k_{\rm B}T=4$, and $\rho^*=6$.
A flat membrane is set along the $xy$ plane with periodic boundary conditions.
The dynamics of the membrane are simulated with a Langevin thermostat as
\begin{equation}\label{eq:dpd}
m \frac{d {\bf v}_{i}}{dt} =
 - \frac{\partial U}{\partial {\bf r}_i} 
 -\zeta{\bf v}_{i} + {\boldsymbol \xi}_i(t),
\end{equation}
where $m$ is the mass of the membrane particle
and $\zeta$ is the friction constant.
The Gaussian white noise ${\boldsymbol {\xi}}_{i}(t)$ 
obeys the fluctuation--dissipation theorem.
The tensionless membrane is used by adjusting the projected membrane area.
On average, each particle interacts with 8 and 19 neighboring particles for $U_{{\rm att},i}$ and $U_{{\rm \alpha},i}$, respectively.
The error bars are estimated from three independent runs.

\begin{figure}
\includegraphics{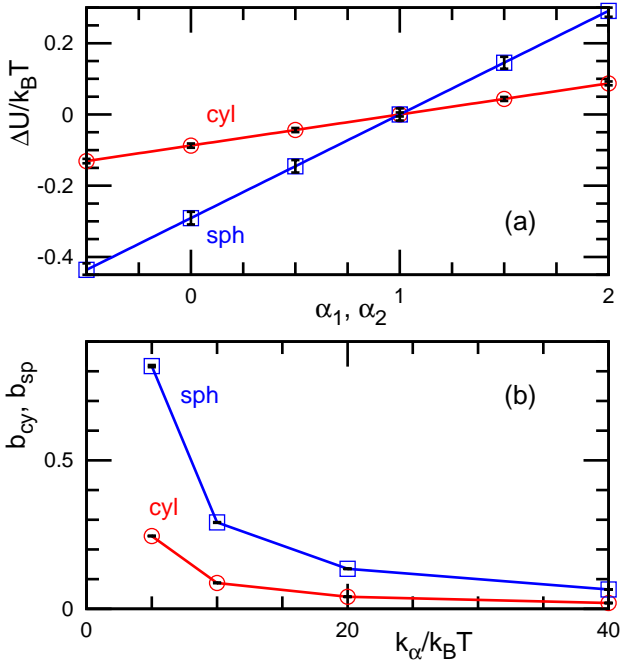}
\caption{
Dependence on $\alpha_1$ and $\alpha_2$ for the meshless membrane model at $N=100$.
The circles and squares represent the data for the cylindrical and spherical deformations, respectively.
(a) Deformation energy compared to that at $\alpha_1=1$ or $\alpha_2=1$ for $k_{\alpha}/k_{\rm B}T=10$.
The solid lines are obtained by a least-squares fit to $\Delta U_{\rm cy}/k_{\rm B}T=b_{\rm cy}\alpha_1 + U_{\rm c0}$ and $\Delta U_{\rm sp}/k_{\rm B}T=b_{\rm sp}\alpha_2 + U_{\rm s0}$.
(b) Slopes of the lines as a function of $k_{\alpha}$.
}
\label{fig:mls2}
\end{figure}

\begin{figure}
\includegraphics{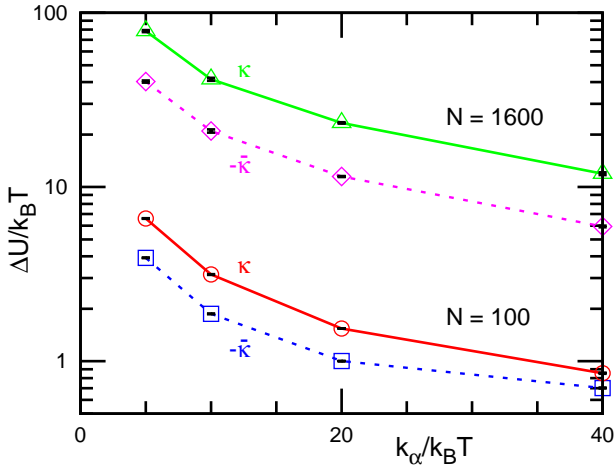}
\caption{
Contributions of the variance terms [the last terms in Eqs.~(\ref{eq:cylf2}) and (\ref{eq:sphf2})] to the deformation energy of the meshless membrane model.
The solid and dashed lines represent the contributions to $\kappa$ and $-\bar{\kappa}$, respectively.
The upper and lower two lines indicate the data at $N=1600$ and $N=100$, respectively.
}
\label{fig:mls3}
\end{figure}

\subsection{Estimation of $\kappa$ and $\bar{\kappa}$}

Figure~\ref{fig:mls1}(b) shows the bending rigidity $\kappa$ estimated by 
the proposed virtual bending method with $\alpha_1=1$ in comparison with the membrane-fluctuation method.
They show excellent agreement in the wide range of $\kappa= 10k_{\rm B}T$--$100k_{\rm B}T$.
The bending rigidity $\kappa$ is linearly dependent on $k_{\alpha}$.
In the membrane-fluctuation method, 
$\kappa$ is calculated from the height fluctuation spectrum $h(q)$ of a flat membrane.
The influence of the short-range particle protrusion is 
removed using extrapolation to the upper-cutoff frequency $q_{\rm {cut}} \to 0$~\cite{shib11}.

The saddle-splay modulus $\bar{\kappa}$ is estimated using the virtual bending method with $\alpha_1=\alpha_2=1$ and 
the membrane closure method described in Appendix~\ref{sec:close} as shown in Fig.~\ref{fig:mls1}(c).
Considering a possible systematic error in the membrane closure method $\Delta\bar{\kappa}/\kappa= \pm 0.1$~\cite{nogu19},
the obtained $\bar{\kappa}$ values agree very well.
In general, for lipid membranes,  $\bar{\kappa}/\kappa \sim -1$~\cite{hu12}.

As described earlier, the proposed method accurately determines the $\kappa$ and $\bar{\kappa}$ values of the meshless membrane.
However, we now survey the parameter dependence in detail.
Figure~\ref{fig:mls2} shows the dependence on the parameters $\alpha_1$ and $\alpha_2$, which are introduced to evaluate
the volume-fluctuation effects.
The work done for cylindrical and spherical deformations ($\kappa$ and $4\kappa+2\bar{\kappa}$)
is slightly dependent on $\alpha_1$ and $\alpha_2$, respectively.
They show a linear dependence for all parameter sets.
The amplitude of the slope decreases with increasing $\alpha_1$ and $\alpha_2$, as shown in Fig~\ref{fig:mls2}(b).
This is due to the suppression of membrane fluctuations at higher bending rigidity at greater values of $k_{\alpha}$.
Therefore, the contribution of these volume-constraint terms is extremely small:
the terms modify $\kappa$ and $\bar{\kappa}$ by less than $0.3k_{\rm B}T$, even if they are removed
as $\alpha_1=\alpha_2=0$.

The contributions of the variance terms are shown in Fig.~\ref{fig:mls3}.
They are the last terms in Eqs.~(\ref{eq:cylf2}) and (\ref{eq:sphf2}) for the cylindrical and spherical deformations,
respectively.
Note that they are independent of $\alpha_1$ and $\alpha_2$.
The contribution can be significant;
it becomes larger than $\kappa$ and $|\bar{\kappa}|$ at low bending rigidity,
although it decreases with increasing $k_{\alpha}$.
Moreover, as the system size is increased from $N=100$ to $N=1600$,
the contribution becomes ten times greater.
Nevertheless, the difference between the obtained values of $\kappa$ and $\bar{\kappa}$ between  $N=100$ and $1600$
is less than $0.6k_{\rm B}T$.
Hence, these large differences in the variance terms are canceled out by those of the other terms.
Since a larger membrane more largely deviates from the energy-minimum state (i.e., the completely flat membrane),
 the second derivatives of the energy increase as well as the variance terms.
If these  variance terms are ignored,
greater values are incorrectly estimated for $\kappa$ and $|\bar{\kappa}|$.
Thus, the variance terms are non-negligible.

The potentials $U_{{\rm att},i}$ and $U_{{\rm \alpha},i}$ are multibody potentials.
For $U_{{\rm att},i}$, we examine three types of center of the deformations:
the geometrical center, the coordinate ${\bf r}_i$ of the center particle,
and the center of the particle pair $({\bf r}_i+{\bf r}_j)/2$ after decomposing the forces to pairwise forces.
For $U_{{\rm \alpha},i}$, we examine two types of center of the deformations:
the center ${\bf r}_{\rm G}$ and the coordinate ${\bf r}_i$ of the center particle.
The influences of these different centers are negligible
 at less than $0.04k_{\rm B}T$.

\section{DPD membrane}\label{sec:DPD}

\subsection{Model}
As an explicit-solvent membrane model, we choose a membrane consisting of two-bead molecules with dissipative particle dynamics (DPD) potentials.
DPD is a coarse-grained molecular simulation method that uses a soft-core repulsive potential and a pairwise Langevin thermostat~\cite{hoog92,espa95,groo97}. 
It has been widely applied to  amphiphilic molecules \cite{vent06,muel06,espa17}.
Here, we use a simple two-bead amphiphilic-molecule model~\cite{naka15}.

The particles interact with each other via a pairwise repulsive potential,
\begin{equation}
U_{\rm DPD}(r_{ij})=  \frac{a_{k\ell}}{2}( r_{ij}/r_{\rm cut} - 1 )^2 \theta(r_{\rm cut}-r_{ij}),
\end{equation}
where $k$ and $\ell$ are particle types.
Here, three types of particles are considered: solvent (w), head (h), and tail (t).
The amphiphilic molecule consists of two particles (h and t) connected by a harmonic bond potential,
\begin{equation}
U_{\rm bond}(r_{i,i+1})=  \frac{b}{2}( r_{i,i+1}/r_{\rm cut} - 1)^2.
\end{equation}
Following Ref.~\onlinecite{naka15}, the parameters
$a_{\rm ww}=a_{\rm hh}=a_{\rm tt}=a_{\rm wh}=100k_{\rm B}T$, $a_{\rm ht}=200k_{\rm B}T$, $a_{\rm wt}=300k_{\rm B}T$, and $b=480k_{\rm B}T$ are used.
A total of 664 amphiphilic molecules and 3856 solvent particles are set in a cubic simulation box with a side length of $12r_{\rm cut}$
under periodic boundary conditions (mean particle density $\rho_{\rm DPD}=3/r_{\rm cut}^3$).
The membrane is normal to the $z$-axis and in a tensionless state in a fluid phase.
The bending rigidity and saddle-splay modulus are calculated as
$\kappa/k_{\rm B}T=18.0 \pm 0.3$ and $\bar{\kappa}/\kappa=-1.06$
using the fluctuation spectrum of a flat membrane and membrane-closure transition, respectively, in Ref.~\onlinecite{naka15}.

The equation of motion for the $i$-th particle with mass $m$ is given by
\begin{equation}\label{eq:dpd}
m \frac{d {\bf v}_{i}}{dt} =
 - \frac{\partial U}{\partial {\bf r}_i} 
 + \sum_{j\not=i} \left\{-w(r_{ij}){\bf v}_{ij}
   \cdot{\bf \hat{r}}_{ij} + 
      {\xi}_{ij}(t)\right\}{\bf \hat{r}}_{ij},
\end{equation}
where the Gaussian white noise ${\xi}_{ij}(t)$ 
obeys the fluctuation--dissipation theorem and
a weight $w(r_{ij})=(1-r_{ij}/r_{\rm cut})$ is used.
The DPD equation~(\ref{eq:dpd}) is discretized by
Shardlow's S1 splitting algorithm~\cite{shar03,nogu07a}.
The results are displayed with the length unit $r_{\rm cut}$.
The error bars are estimated from three or ten independent runs.

\begin{figure}
\includegraphics{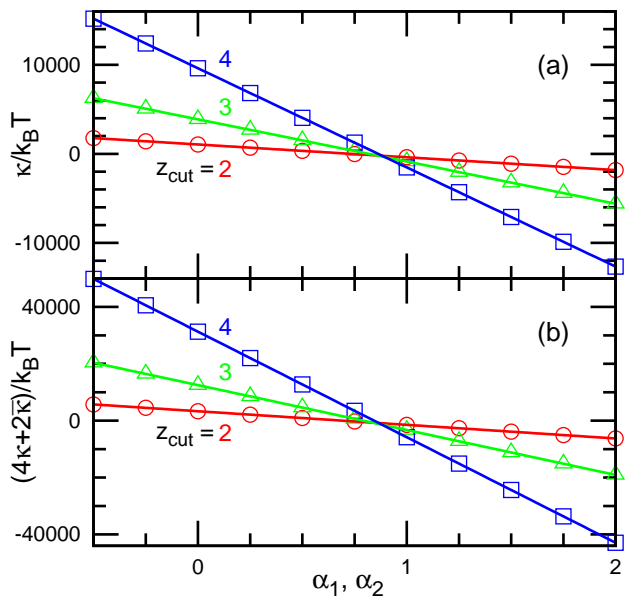}
\caption{
Dependence on $\alpha_1$ and $\alpha_2$ for the simple DPD fluid at $a_{\rm ww}=100k_{\rm B}T$
for the (a) cylindrical and (b) spherical deformation.
The circles, triangles, and squares represent data with cutoff $z_{\rm cut}=2$, $3$,  and $4$, respectively.
The error bars are smaller than the line thickness.
}
\label{fig:dpdw}
\end{figure}

\subsection{Simple DPD Fluid}

Before discussing the membrane,
we consider a simple DPD fluid consisting of a single type (w) of particles.
The density $\rho_{\rm DPD}=3/r_{\rm cut}^3$ is employed.
The cubic simulation box with a side length of $12r_{\rm cut}$ is used unless otherwise specified.
Because it is in a fluid state, the simple DPD fluid can freely change its shape while maintaining a constant volume;
hence, it exhibits  zero surface tension and zero bending rigidity.

Figure \ref{fig:dpdw} shows the free-energy changes for the cylindrical and spherical deformations at $a_{\rm ww}=100k_{\rm B}T$
from Eqs.~(\ref{eq:cylf2}) and (\ref{eq:sphf2}), respectively.
The virtual deformations of the region $|z|<z_{\rm cut}$ are calculated.
When an interacting particle pair crosses the boundary ($\pm z_{\rm cut}$),
the first and second derivatives are linearly divided into two regions with a ratio of $|(z_{\rm cut}-z_i)/z_{ji}|$.
Note that it is not sensitively dependent on the division methods; when $|(z_{\rm cut}^2-z_i^2)/(z_{j}^2-z_i^2)|$ is used instead for the second derivatives, 
only negligible differences are found.
The energy changes exhibit a linear dependence on the parameters $\alpha_1$ and $\alpha_2$, as shown in Figs.~\ref{fig:dpdw}(a) and (b), respectively,
and can vanish at $\alpha_{1}$ and $\alpha_{2}= 0.7$--$0.9$: $\alpha_1=0.7331 \pm 0.0003$, $0.8185 \pm 0.0002$, and $0.8632 \pm 0.0004$, and
$\alpha_2=0.6945\pm 0.0003$, $0.7923\pm 0.0002$, and $0.8435\pm 0.0004$ for $z_{\rm cut}=2$, $3$, and $4$, respectively.
Hence, the values at vanishing energy changes slightly increase with an increase in the cutoff $z_{\rm cut}$.
As $z_{\rm cut}$ increases, the magnitudes of the slopes increase.
We also simulate the fluid for two rectangular simulation boxes with different lateral lengths as $6r_{\rm cut}\times 6r_{\rm cut}\times 12r_{\rm cut}$ 
and $24r_{\rm cut}\times 24r_{\rm cut}\times 12r_{\rm cut}$.
The differences of the obtained values of $\kappa$, $\bar{\kappa}$, and also the variance terms are in the order of the statistical errors,
so that simple fluids have no system size effects.

Moreover, we simulate the DPD fluid at $a_{\rm ww}/k_{\rm B}T=25$, $50$, and $150$.
A similar dependence on $z_{\rm cut}$ is obtained. 
The magnitudes of the slopes linearly increase  with increasing $a_{\rm ww}$.
By contrast, the vanishing values of $\alpha_{1}$ and $\alpha_{2}$  decrease  approximately linearly with increasing $a_{\rm ww}$ 
and are in the range of $0.5$--$1$.
Thus, the virtual bending method cannot calculate $\kappa$ and $\bar{\kappa}$ without adjustments to $\alpha_{1}$ and $\alpha_{2}$.

\begin{figure}
\includegraphics{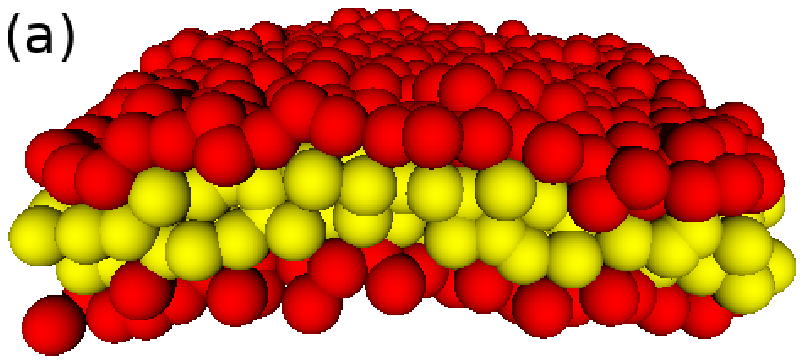}
\includegraphics{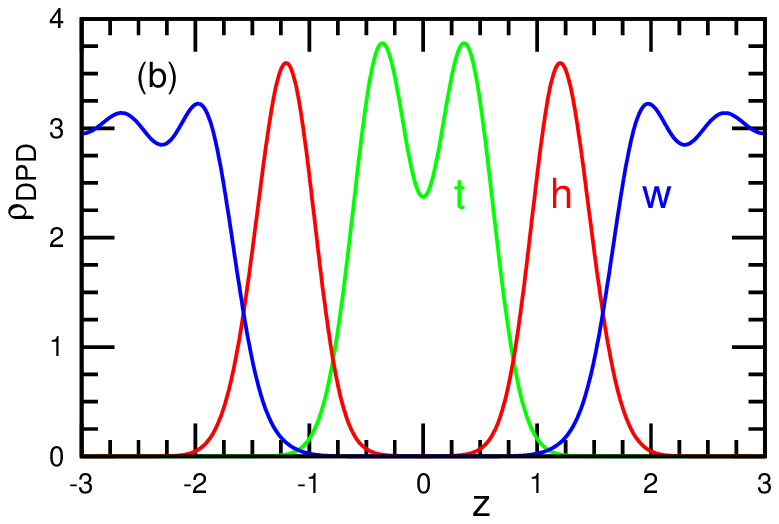}
\caption{
Membrane of the two-bead DPD model.
(a) Snapshot of the membrane. The red and yellow spheres represent the head (h) and tail (t) particles, respectively.
The solvent (w) particles are not shown for clarity.
(b) Vertical profile of the number density $\rho_{\rm DPD}$ of the head (h), tail (t), and solvent (w) particles  along the $z$-axis. 
The origin of the $z$-axis is set at the center of the mass of the membrane.
The error bars are smaller than the line thickness.
}
\label{fig:dpd1}
\end{figure}

\begin{figure}
\includegraphics{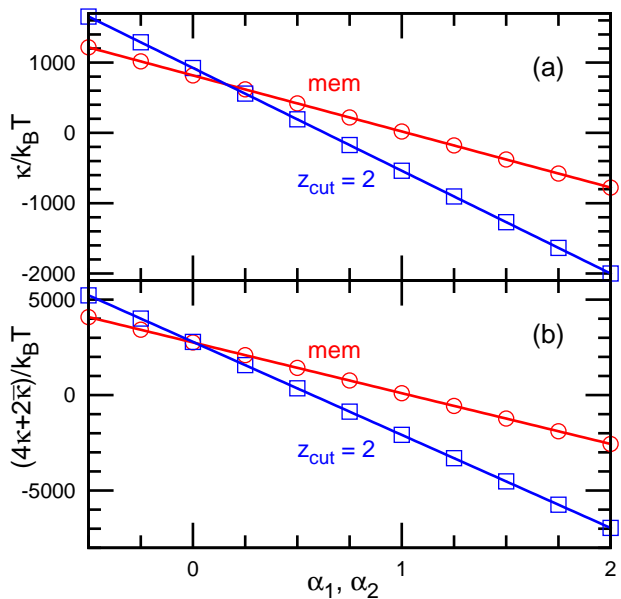}
\caption{
Dependence on $\alpha_1$ and $\alpha_2$ of the two-bead DPD model
for the (a) cylindrical and (b) spherical deformation.
The circles and squares represent data with cutoff at the membrane--solvent interface and $z_{\rm cut}=2$, respectively.
The error bars are smaller than the line thickness.
}
\label{fig:dpd2}
\end{figure}

\subsection{Membrane}

Next, we consider a membrane consisting of two-bead DPD amphiphiles.
As shown in Fig.~\ref{fig:dpd1}, the membrane forms a clear bilayer structure.
The free-energy changes for the cylindrical and spherical deformations
similarly exhibit a linear dependence on $\alpha_1$ and $\alpha_2$ (Fig.~\ref{fig:dpd2}), 
as obtained in the DPD fluid.
When the energy changes are cutoff at the membrane interface,
the interactions between the solvent particles are not considered,
and the interactions between the solvent and membrane particles are half considered.
The expected values [$\kappa/k_{\rm B}T=18$ and $(4\kappa+2\bar{\kappa})/k_{\rm B}T=36$]
are obtained at $\alpha_1=1.00\pm 0.03$ and $\alpha_2=1.02\pm 0.03$ for the membrane--solvent interface cutoff,
and at $\alpha_1=0.62\pm 0.02$ and $\alpha_2=0.56\pm 0.02$ for $z_{\rm cut}=2$, respectively.
Thus, $\kappa$ and $\bar{\kappa}$ are estimated at the local-volume conservation condition ($\alpha_1=\alpha_2=1$)
in the case of the membrane--solvent interface cutoff.
However, since they largely depend on $\alpha_1$ and $\alpha_2$,
it is very difficult to precisely estimate  $\kappa$ and $\bar{\kappa}$ using the present scheme.
An extension of the present method to account for the volume fluctuations is required 
to apply it to membranes in explicit solvents.

\section{Summary and Discussions}\label{sec:sum}

We have proposed a calculation method for $\kappa$, $\bar{\kappa}$, and $C_0$ 
of a fluid membrane.
The free-energy changes caused by the cylindrical and spherical deformations
are rigorously derived.
The first derivative is a discrete form of the first moment of the stress profile.
We also clarified that the first moment is independent of the choice of the angular-momentum conserving force-decomposition.
The second derivatives involve the variance term of the first derivative to express the thermal-fluctuation effects 
that is not considered in the previous stress-profile method.

For the meshless membrane model, $\kappa$ and $\bar{\kappa}$ can be accurately calculated from the second derivatives.
This requires a significantly smaller membrane size than other methods.
The effects of the volume fluctuations are negligible.
By contrast, for the DPD membrane model, excessively large volume-fluctuation effects are found, requiring
 an extension to consider these effects for a reasonable estimation of  $\kappa$ and $\bar{\kappa}$.
Thus, the proposed method works very well for a thin membrane in a vacuum or an implicit-solvent.
However, it requires a further extension for a thick membrane or an explicit-solvent condition.
The magnitude of the volume-fluctuation effects can be estimated from the dependence on the volume parameters $\alpha_1$ and $\alpha_2$.

In general, 
macroscopic quantities such as $\kappa$ and $\bar{\kappa}$ should not depend on an arbitrary choice of the local force fields.
Here, the first derivative for  $\kappa C_0$ is independent of the choices of the force decomposition or the lateral center position of deformation.
The second derivatives for $\kappa$ and $\bar{\kappa}$ are dependent not on the force decomposition but
on the center position.
This is because the second-order deformation depends on the center position.
However, it is found that this dependence is negligible when the center is chosen in a reasonable manner.
If one pursues the rigid uniqueness, the center can be likely determined as the position minimizing the deformation energy.
Because the second moment of the stress profile is significantly affected by the choices of the force decomposition,
the stress-profile method is unsuitable for calculating $\bar{\kappa}$ even if the fluctuation effects are included.
However, for deterministic continuum simulations,
the stress-profile method may be applicable for $\bar{\kappa}$ estimation
because thermal fluctuations are not accounted for and the local stress is explicitly defined.

The present virtual deformation method can be applied to different types of material deformation, such as shear and twisting.
For the first derivative of the deformation, the present procedure can be straightforwardly applied to other systems.
However, a careful examination of the fluctuation effects is required for the second derivatives.

\begin{acknowledgments}
This work was supported by JSPS KAKENHI Grant Number JP17K05607.
The simulations were partially
carried out by HPE SGI 8600 at the ISSP Supercomputer Center, University of Tokyo. 
\end{acknowledgments}

\begin{appendix}

\section{Uncertainty of Force Decomposition}\label{sec:ccfd}

Although the global stress is uniquely determined in molecular simulations using 
the virial calculation,
a local stress field depends on the choice of force propagation pathway~\cite{adma10,naka16}.
When forces are assumed to propagate along straight lines between interacting particles,
the forces of $U_{k_n}$ are decomposed into central forces of the form ${\bf f}_i= \sum_j^{n-1} f_{ij}\hat{\bf r}_{ij}$. 
As the forces have $2n-3$ and $3(n-2)$ degrees of the freedom in the 2D and 3D spaces,
the values of $n(n-1)/2$ force pairs are not uniquely determined for $n>3$ and $n>4$, respectively.

When the force pairs are represented by an ($n(n-1)/2$)-dimensional vector ${\boldsymbol {\Psi}} = (f_{12}, ... f_{ij},...)$,
the available force pairs are expressed as~\cite{torr16,adma16}
\begin{equation}
 \label{eq:cfdpair}
{\boldsymbol{\Psi}} = {\boldsymbol{\Psi}}_0 + \sum_{\ell }^{n_{\rm rest}} a_\ell {\boldsymbol{\Psi}}_\ell
\end{equation}
where $n_{\rm rest}= n(n-1)/2 - (2n-3)$ and $n_{\rm rest}= n(n-1)/2 - 3(n-2)$ in the 2D and 3D spaces, respectively,
and $a_\ell$ is an arbitrary real number.
${\boldsymbol{\Psi}}_\ell$ gives no forces on the particles (${\bf f}_{i,\ell}= \sum_j^{n-1} f_{ij,\ell}\hat{\bf r}_{ij}= 0$)
and can be determined by using Cayley--Menger determinants for a $k$D volume, where $k>3$ and $k>4$ in the 2D and 3D spaces, respectively.
${\boldsymbol{\Psi}}_0$ is taken orthogonally to the other vectors as ${\boldsymbol{\Psi}}_0\cdot {\boldsymbol{\Psi}}_\ell=0$,
so that ${\boldsymbol{\Psi}}_0$ is uniquely determined.
${\boldsymbol{\Psi}}_0$ is called a covariant~\cite{torr16} or irrotational~\cite{adma16} component.
In the covariant CFD, ${\boldsymbol{\Psi}}_0$ is used as the unique force decomposition~\cite{torr16}.
In other words, the covariant CFD chooses the decomposition that yields the minimum of ${\boldsymbol {\Psi}}^2=\sum_{ij} f_{ij}^2$.

Let us consider an $n$-body potential $U_{\rm ps}$ that can be expressed by the sum of pairwise potentials in a certain limit,
 $U_{\rm ps}= \sum_{k} U_{{\rm pair},k}(r_{ij})$, for $n>4$.
In this limit,  the forces should be decomposed into these pairwise forces $f_{ij}= - \partial U_{{\rm pair},k}/\partial r_{ij}$.
However,  the covariant CFD or any other decomposition based only on the forces ${\bf f}_i$ cannot 
predict the correct decomposition because the forces do not have sufficient information for it.
In specific cases such as the aforementioned pairwise potentials, 
the force decomposition may be chosen for a physical or mathematical reason.
However, the decomposition is not uniquely determined in general.
A similar non-uniqueness problem appears in the discretization of the Navier--Stokes equation (NSE)~\cite{nogu19b}.
Since the form of the stress field is known in the NSE, the correct discretization is obtained by following the stress field.

\begin{figure}
\includegraphics{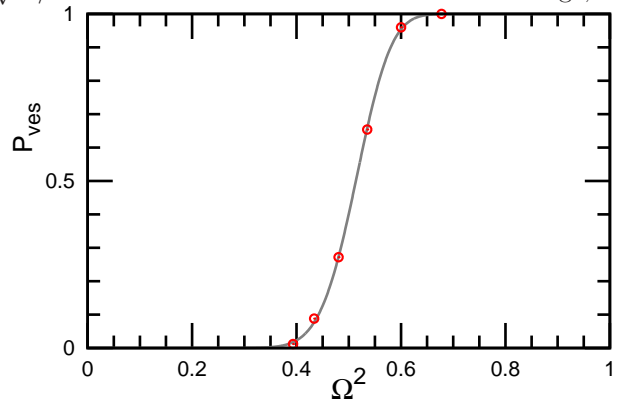}
\caption{
Closing probability $P_{\rm {ves}}$ of a membrane patch for the meshless membrane model 
at $N=1500$ and $k_{\alpha}=20k_{\rm B}T$.
The circles represent the simulation data.
Each probability is calculated from 500 samples.
The solid gray lines are obtained through fitting to Eq.~(\ref{eq:close}).
}
\label{fig:close}
\end{figure}

\section{Membrane Closure}\label{sec:close}

The saddle-splay modulus $\bar{\kappa}$ can be calculated from the transition between a circular membrane patch and a spherical vesicle~\cite{hu12}.
When a spherical cap with the curvature radius $R$ is assumed to be in an intermediate state,
the excess energy, $\Delta E$, of the membrane with respect to the flat circular membrane patch is given by~\cite{from83}
\begin{equation}
 \label{eq:chp3eq1}
 \Delta E (\Omega^{2},\tilde{\Gamma}) = 4\pi(2\kappa + \bar{\kappa})\big[\Omega^{2} + \tilde{\Gamma}(\sqrt{1 - \Omega^{2}} - 1)\big],
\end{equation}
where
$\tilde{\Gamma} = \Gamma R_{\rm {ves}}/(2 \kappa + \bar{\kappa})$, $\Omega=R_{\rm {ves}}/R$, and $R_{\rm {ves}} = \sqrt{A/4\pi}$.
$\Gamma$ is the line tension of the membrane edge, and $A$ is the membrane area.
The normalized curvature $\Omega$ is an order parameter.
At $\tilde{\Gamma}=1$, the flat patch ($\Omega=0$) and vesicle ($\Omega=1$) have the same energy and, for
$0<\tilde{\Gamma}<2$, a free-energy barrier exists at $0<\Omega<1$.

The free-energy barrier can be determined by collecting samples in which
pre-curved membranes change into open disks or closed vesicles. 
The probability $P(\Omega^{2})$ of this
change occurring is derived for the initial $\Omega$
as \cite{hu12} 
\begin{equation}
 P_{\rm {ves}}(\Omega^{2}) = \dfrac{ \int_{0}^{\Omega^{2}} du\ \exp\left(\frac{\Delta
  E(u,\tilde{\Gamma})}{\tilde{D}}\right)  }{ \int_{0}^{1} du\ \exp\left(\frac{\Delta E(u,\tilde{\Gamma})}{\tilde{D}}\right)  }.
\label{eq:close}
\end{equation}
The parameters $\tilde{D}$ and $\tilde{\Gamma}$ are obtained by fitting this function to the simulation data;
 $\bar{\kappa}$ is determined from the value of $\tilde{\Gamma}$ with the separately calculated $\kappa$ and $\Gamma$.
This method is first applied to a solvent-free molecular model~\cite{hu12} and 
later to the MARTINI model~\cite{hu13}, the two-bead DPD-molecule model (Sec.~\ref{sec:DPD})~\cite{naka15},
and a spin meshless membrane model~\cite{nogu19}.
Here, we calculated  $\bar{\kappa}$ of the moving-least-squares meshless membrane as employed in Sec.~\ref{sec:meshless}.
The closing probability, $P_{\rm {ves}}$ is fit to Eq.~(\ref{eq:close}) very well as shown in Fig.~\ref{fig:close}.
Hence,  $\bar{\kappa}/\kappa =-1.04\pm 0.01$ is obtained from ${\kappa}/k_{\rm B}T=44.1\pm 0.6$ and $\Gamma\sigma/k_{\rm B}T=4.46\pm 0.03$.

\end{appendix}

\end{document}